\documentclass[10pt]{article}
\usepackage{times}
\usepackage{fullpage}
\usepackage{amsmath,amsthm}
\usepackage{latexsym}
\usepackage{graphicx}
\usepackage{amssymb}
\usepackage{color}
\usepackage{xspace}
\usepackage{multicol,multirow}
\usepackage{algorithm, algorithmic}

\graphicspath{{.}}

\newcommand{\opt}{\ensuremath{\text{\it OPT}}}
\newcommand{\CA}{\ensuremath{{\cal A}}}
\newcommand{\CR}{\ensuremath{{\cal R}}}

\newcommand{\CP}{\ensuremath{{\cal P}}}
\newcommand{\CS}{\ensuremath{{\cal S}}}
\newcommand{\CX}{\ensuremath{{\cal X}}}

\newcommand{\CF}{\ensuremath{{\cal F}}}

\newcommand{\CT}{\ensuremath{{\cal T}}}

\newcommand{\ignore}[1]{}

\newcommand{\etal}{et al.\xspace}

\newcommand{\sensorcover}{{\sc Sensor Cover}\xspace}
\newcommand{\intervalcover}{{\sc Restricted Strip Cover}\xspace}
\newcommand{\rectanglecover}{{\sc Cube Cover}\xspace}
\newcommand{\odpname}{{\intervalcover}\xspace}
\newcommand{\odspname}{{\sc RSC}\xspace}
\newcommand{\tdpname}{{\rectanglecover}\xspace}

\newcommand{\aspname}{{\sc Sensor Cover}\xspace}
\newcommand{\dsa}{{\sc Dynamic Storage Allocation}\xspace}
\newcommand{\sdsa}{{\sc DSA}\xspace}
\newcommand{\tpp}{{\sc 3-Partition}\xspace}
\newcommand{\scp}{{\sc Set Cover Packing}\xspace}

\newcommand{\naesat}{{\sc NAE-3SAT}\xspace}
\newcommand{\nib}[1]{\noindent{\bf #1}}

\newtheorem{theorem}{Theorem}[section]
\newtheorem{lemma}[theorem]{Lemma}

\newtheorem{corollary}[theorem]{Corollary}
\newtheorem{proposition}[theorem]{Proposition}

\newtheorem*{problem}{Problem}

\newcommand{\hmin}{d_{\min}}
\newcommand{\hmax}{d_{\max}}
\newcommand{\eps}{\epsilon}
\newcommand{\ith}{^\text{th}}

\newcommand{\newpar}[1]{\medskip\noindent{\bf #1}}

\makeatletter
\newcommand{\poly}{\mathop{\operator@font poly}}
\newcommand{\argmax}{\mathop{\operator@font argmax}}
\makeatother

\title{Restricted Strip Covering and the Sensor Cover Problem}
\author{Adam L.~Buchsbaum\thanks{AT\&T Labs, {\sf alb@research.att.com}.}
\and
Alon Efrat \thanks{Computer Science Department, University of Arizona,
  {\sf alon@cs.arizona.edu}.} 
\and
Shaili Jain \thanks{Division of Engineering and Applied Sciences, Harvard University, {\sf shailij@eecs.harvard.edu}.
  Part of the work was done while visiting AT\&T Shannon Labs.  Supported by the AT\&T Labs Fellowship Program.}
\and
Suresh Venkatasubramanian \thanks{AT\&T Labs, {\sf suresh@research.att.com}.}
\and
Ke Yi \thanks{Computer Science Department, Duke University, {\sf
    yike@cs.duke.edu}.  Part of the work was done while visiting AT\&T Shannon Labs.}
}

\begin{document}

\maketitle
\begin{abstract}
Suppose we are given a set of objects that cover a region and a duration
associated with each object.  Viewing the objects as jobs, can we schedule
their beginning times to maximize the length of time that the original
region remains covered?  We call this problem the \textsc{Sensor Cover
  Problem}. It arises in the context of covering a region with sensors.
For example, suppose you wish to monitor activity along a fence (interval)
by sensors placed at various fixed locations.  Each sensor has a range
(also an interval) and limited battery life.  The problem is then to
schedule when to turn on the sensors so that the fence is fully monitored
for as long as possible.

This one dimensional problem involves intervals on the real line.
Associating a duration to each yields a set of rectangles in space and
time, each specified by a pair of fixed horizontal endpoints and a height.
The objective is to assign a bottom position to each rectangle (by moving
them up or down) so as to maximize the height at which the spanning
interval is fully covered.  We call this one dimensional problem {\sc
  Restricted Strip Covering}.  If we replace the covering constraint by a
packing constraint (rectangles may not overlap, and the goal is to
minimize the highest point covered), then the problem is identical to
\dsa, a well-studied scheduling problem, which is in turn a restricted case
of the well known problem \textsc{Strip Packing}.

We present a collection of algorithms for {\sc Restricted Strip Covering}.
We show that the problem is NP-hard and present an $O(\log \log
n)$-approximation algorithm.  We also present better approximation or
exact algorithms for some special cases.  For the general {\sc Sensor
  Cover Problem}, we distinguish between cases in which elements have
uniform or variable durations.  The results depend on the structure of the
region to be covered: We give a polynomial-time, exact algorithm for the
uniform-duration case of {\sc Restricted Strip Covering} but prove that
the uniform-duration case for higher-dimensional regions is NP-hard.
Finally, we consider regions that are arbitrary sets, and we present an
$O(\log n)$-approximation algorithm for the most general case.
\end{abstract}

\section{Introduction}

Sensors are small, low-cost devices that can be placed in a region to
monitor local conditions.  Distributed sensor networks have become
increasingly more popular as advances in MEMS and fabrication allow for
such systems that can perform sensing and communication. How sensors
communicate is a well-studied problem.  Our main interest is:  Once a
sensor network has been established, how can we maximize the lifetime of
the network?  It is clear that the limited battery capacities of sensors
is a key constraint in maximizing the lifetime of a network. Additionally,
research shows that partitioning the sensors into covers and iterating
through them in a round-robin fashion increases the lifetime of the
network~\cite{AGP04,DVCS04,PH03,SP01}.

\newpar{Definitions.}
Let $\CS = \{s_1,\dots, s_n\}$ be a set of sensors.  Each sensor $s$ can be
viewed as a point in some space with an associated region $R(s)$ of
coverage.  For every point $x\in R(s)$, $s$ is said to be {\em live at
  $x$}.  Let $U$ be the region to be covered by the sensors. $U$ is
\emph{covered} by a collection ${\CR} \subseteq \CS$ of sensors if $U \subseteq
\bigcup_{s \in {\CR}} R(s)$. We often refer to ${\CR}$ as a \emph{feasible
  cover}. 
Every sensor $s\in\CS$ can be {\em active} for a finite {\em duration}
$d(s)$.  Let $\hmin = \min_{s\in\CS} d(s)$, and $\hmax = \max_{s\in\CS}
d(s)$.  

\begin{problem}[\sensorcover]
Compute a {\em schedule} $S$
of maximum {\em duration} $T$,
in which each sensor $s\in\CS$ is assigned a {\em start time} $t(s)\ge 0$,
such that any $x\in U$ is covered by some active sensor at all times $0\le t
< T$.
That is, for all $x\in U$ and $0\leq t<T$, there is some $s\in \CS$ with 
$x \in R(s)$ and $t(s) \le t < t(s)+d(s)$.
\end{problem}

A sensor is \emph{redundant} in a schedule $S$ if it can be
removed without decreasing the duration of $S$.  A schedule with
no redundant sensors is a \emph{minimal} schedule.  By removing the
redundant sensors, any optimal schedule can be converted to a minimal
schedule of the same total duration.  Therefore it suffices to 
consider only minimal schedules,
which may not utilize all sensors.  As a convention, we set $t(s) =
\infty$ if $s$ is unused.

Prior work on the \aspname problem has focused solely on the case where the
regions $R(s)$ are arbitrary subsets of $U$ and the durations are all
identical. This assumption yields a \emph{packing} constraint, and the
problem reduces to partitioning the set of sensors into a maximum number
of valid covers. This problem is known as \scp and is $\ln n$-hard to
approximate, with a matching upper bound \cite{FHK00}. 

In practice though, these assumptions appear overly constraining. Sensors
will have arbitrary durations and typically define geometric regions of
coverage: intervals, rectangles, disks, etc. In this paper, we will
consider these classes of problems.  In the \intervalcover problem, all
the $R(s)$'s are intervals in one dimension.  In this case the problem is
equivalent to sliding axis-parallel rectangles vertically to cover a
rectangular region of maximum height.  Thus, \odpname resembles the \dsa
(\sdsa) problem \cite{GJ79}, albeit in a dual-like fashion.  In the
\rectanglecover problem, the $R(s)$'s are axis-parallel rectangles, and
the problem is akin to sliding cubes vertically in the $z$-dimension.  We
also consider \sensorcover, when the $R(s)$'s are arbitrary subsets of a
finite set $U$ of size $|U| = O(n)$, with varying durations (in
contrast to \scp).

In general, a schedule may activate and deactivate a sensor more than once.
We call this a {\em preemptive schedule}.  A \emph{non-preemptive schedule}
is a schedule in which each sensor is activated at most once.  In this
paper we only consider the non-preemptive problem.  We have some
preliminary results for the preemptive case, but more research is needed to
gain a better understanding of the differences.

\newpar{Our results.}
We show that most variants of \sensorcover are NP-hard,
and we study approximation algorithms.  For any point $x\in U$, let
$L(x) = \sum_{s \in\CS, s \text{ live at } x} d(s) $ be the {\em load at}
$x$ .  Define the overall {\em load} $L=\min_{x} L(x)$.  We write $L_X$
(rsp., $L_X(x)$) for the load of any subset $X$ of sensors (rsp., at $x$).
Letting $\opt$ denote the duration of an optimal schedule, a trivial upper
bound is $\opt\leq L$.  All our approximation ratios are with
respect to $L$.  That $\opt \le L$ allows the assumption that
$\hmax\leq L$, because durations exceeding $L$ contribute to neither load
nor $\opt$. 

Table~\ref{tab:results} summarizes our results. 
The most interesting case
is the \odpname problem, for which we give an $O(\log\log
n)$-approximation algorithm. 
\begin{table}[t]
\begin{center}
\caption{Summary of results.}
\begin{tabular}{c|c|c}
  Shape of sensor & Uniform duration & Variable duration \\ \hline\hline
  Intervals &          exact in P    & NP-hard, $O(\log\log n)$-approx. \\ \hline
Rectangles, Disks, \ldots & NP-hard, $O(\ln
  (n/L))$-approx. & NP-hard, $O(\log n)$-approx.\\\hline
\multirow{2}*{Arbitrary sets} & $\log n$-hard to approx.,   & $\log n$-hard to approx.,\\
 & $O(\log n)$-approx.~\cite{FHK00}& $O(\log  n)$-approx.\\
\end{tabular}
\label{tab:results}
\end{center}
\vspace{-5ex}
\end{table}
After reviewing some related work, we discuss \odpname in
Section \ref{sec:interval}, \tdpname in Section \ref{sec:rectangle}, and
the general \aspname problem in Section \ref{sec:kcover}.

\newpar{Related Work.} \label{sec:rel}
\scp was studied by Feige \etal~\cite{FHK00}. They considered the
{\sc Domatic Number} problem, where the goal is to maximize the number of
disjoint dominating sets on the set of vertices of a graph.  A
\emph{dominating set} in a graph $G = (V,E)$ is a set $V'\subset V$ of vertices 
such that every $v \in V$ is either contained in $V'$ or
has a neighbor in $V'$.  Feige \etal show for every $\varepsilon > 0$, the
{\sc Domatic Number} problem is hard to approximate within a factor of
$(1-\varepsilon)\ln |V|$.  In the proof of their hardness result, Feige
\etal use the \scp problem, whose goal is to maximize the number of
disjoint set covers, given a set of subsets $S$ of a base set $U$.  Note
that the \scp problem is a combinatorial version of our problem,
with each subset being a region, and each (sensor) region having unit
duration.
Feige \etal also give a randomized $\ln
n$-approximation algorithm, which they derandomize.  Another key feature of
this work is that it showed the first maximization problem proved to be
approximable within polylogarithmic factors but no better.

The practical motivations for studying this problem
have led to the development of numerous heuristics.  Slijepcevic and
Potkonjak~\cite{SP01} introduce the {\sc Set K-Cover} problem, where they
are given a set of subsets of a base set and an integer $k$ and
ask if it is possible to construct at least $k$ disjoint set covers. They
re-prove an NP-hardness result for this problem, probably unaware of the
Feige \etal result.  They also present a heuristic that selects mutually
exclusive sets of nodes, where each set completely covers the desired
region.

Perillo and Heinzelman~\cite{PH03} study a variation of this problem, where
they want to maximize the lifetime of a multi-mode sensor network. 
They compute all possible feasible covers and then translate their
problem instance into a graph.  Each sensor and feasible cover
becomes a node.  Sensors are connected to a feasible cover if they are
contained in that feasible cover.  They use linear programming to model
additional energy constraints and solve the maximum flow problem on this
graph.  Although they solve the problem optimally, their solution can be exponential in
the problem instance. 
Dasika \etal\cite{DVCS04} also compute all possible feasible covers and
develop heuristics for switching between these covers in order to maximize
the lifetime of their sensor network.

Abrams \etal \cite{AGP04} study a variation of the problem where they are given a
collection of subsets of a base set and a
positive integer $k \geq 2$. Their goal is to partition the subsets into $k$
covers, where the area of coverage, defined as the cardinality of a set, is
maximized across all $k$ covers. 
They give three approximation algorithms for this problem:
a randomized algorithm, a
distributed greedy algorithm, and a centralized greedy algorithm.  Their
randomized algorithm partitions sensors within $1-\frac{1}{e}$ of the
optimal solution.  Their distributed greedy algorithm gives a
$\frac{1}{2}$-approximation ratio.  Their centralized greedy algorithm
achieves an approximation factor of $1-\frac{1}{e}$.  They also prove a
$\frac{15}{16}$-hardness result for their problem.

To our knowledge, the \intervalcover problem has not previously been
considered.  Some of the closely related problems are
well studied, however. If we replace the covering constraint by a packing
constraint (rectangles may not overlap, and the goal is to minimize
the height of the highest point covered), then the problem is
the same as \dsa~\cite[Problem SR2]{GJ79}, for which there is a
$(2+\eps)$-approximation~\cite{dsa:b+04}. If we further allow rectangles
to move both vertically and horizontally, then the problem becomes
\textsc{Strip Packing}, which has a
$(1+\eps)$-approximation (up to an additive term) \cite{strip:kr00}.


\section{Restricted Strip Cover}
\label{sec:interval}

Consider an instance $\CS$ of \odpname (\odspname).  For ease of
presentation, we define $R(s)$ as a semi-closed interval $[\ell(s),r(s))$
for each $s \in \CS$.  
We assume without loss of generality that all interval coordinates are
non-negative integers in the range $[0,2n-1]$ and $U = \bigcup_{s\in \CS} R(s)
= [0, 2n-1)$, because there are at most $2n$ distinct interval endpoints of
sensors.
It is convenient to view scheduled sensors as
semi-closed rectangles in the plane, with their intervals along the
$x$-axis and durations along the $y$-axis.  Thus a valid schedule $S$
of duration $T$ is
one in which any point $(x,y)$ in the sub-plane $U\times[0,T)$ is covered by some sensor $s$;
i.e., $\ell(s) \le x < r(s)$ and $t(s)\le y < t(s)+d(s)$.  The
problem is equivalent to sliding axis-parallel rectangles vertically to
cover a rectangular region of maximum height.  Therefore, in this section we
use the terms ``sensor'' and ``rectangle'' interchangeably.  We say two or
more rectangles {\em overlap} if they cover some common point.  When
discussing multiple schedules, we write $t_S(s)$ to denote the start time
of $s$ in some schedule $S$; otherwise we omit the subscript.

We also assume that all durations
are positive integers.  We define {\em level} $j$ of any schedule
to be the horizontal slice of sensors that cover points at $y$-coordinates
in $[j-1,j)$.  Let $S$ be some schedule of \CS.  A {\em gap} is a point $p$
such that no sensor covers $p$.  For any $i \in U$, define $M(S,i)$ to be the
greatest $y$-coordinate $j$ such that no gap exists below $j$ at $i$; i.e.,
$ M(S,i) = \max\{ j : \forall j' < j, \exists s\in S,\ s \text{ covers }
(i,j') \}$. Then the {\em duration} of $S$ is $M(S) = \min_i M(S,i)$.

Our main result for \odspname is an $O(\log\log n)$-approximation.
En-route we obtain better results for some special cases: 
(i) a
simple, exact algorithm if all sensors have the same duration
(Section~\ref{sec:unif-durat-sens});
(ii) an exact, dynamic programming
algorithm, which runs in $\poly(n)$ time if $L=O(\log n / \log\log n)$
(Section~\ref{sec:dynam-progr-solut});
and 
(iii)
$1/(1+\epsilon)$-approximations if $L=O(\hmin \log n/\log\log n)$ or
$L=\Omega(\hmax \log n \cdot (1/\epsilon)^4 \min\{1/\epsilon,
\log(\hmax/\hmin)\})$ (Section~\ref{sec:appr-algor-via}).

Our main technique builds on the method outlined by Buchsbaum
\etal \cite{dsa:b+04} in their algorithm for \dsa. Their approach does not
apply directly, because the covering constraint poses different
challenges to solving \odspname
than does the packing constraint to \dsa. We will adapt parts of their method
to obtain our bounds. 

\subsection{Uniform-Duration Sensors}
\label{sec:unif-durat-sens}
If all sensors have the same duration, a simple greedy algorithm gives an exact
solution of duration $L$.  Define $\CS_i = \{ s\in \CS: s \text{ is live at
} i\}$.  Assume by scaling that all sensors have unit duration.  We proceed
left-to-right, starting at $i=0$ and constructing a schedule $S$ while
maintaining the following invariants after scheduling sensors in $\CS_i$:
(i) no sensors overlap at any $x$-coordinate $\geq i$, and (ii) $M(S,i)=L$.

When $i=0$, select any $L$ sensors that are live at $0$, and schedule them
without overlap, establishing the initial invariants.  Assuming the
invariants are true at $i$,  schedule $\CS_{i+1}$ as follows.  If there
are no gaps at $i+1$, we are done, as the invariants extend to $i+1$.
Otherwise, assume there are $k>0$ unit-duration gaps at $i+1$.  The
invariants imply that at least $k$ sensors in $\CS_{i+1}$ remain
unscheduled,
which can be used to fill the gaps while preserving the invariants.

\subsection{Hardness Results}
\label{sec:icnp}

When sensors have variable durations, the problem becomes NP-hard.  To
prove this, we exploit an identity to \dsa in a special case.  An instance
of \sdsa is like one of \odspname, except that the instance load is defined
as the maximum load at any $x$-coordinate, and the goal is to schedule the
sensors ({\em jobs}, in \dsa parlance) without overlap so as to minimize
the makespan.
If the load is equal for all $x$-coordinates, then $\opt=L$ for either
problem implies a schedule that is a solid rectangle of height $\opt=L$.
For \odspname, it further implies that the schedule is non-overlapping; for
\sdsa, this implication is redundant.

Stockmeyer proves that determining if there exists a solution to \sdsa of
a given makespan is NP-complete \cite[Problem SR2]{GJ79}.  His proof reduces
an instance of \tpp to an instance of \sdsa with uniform load and,
in fact, all durations in $\{1,2\}$, such that $\opt=L$ if and
only if there is a solution to the \tpp instance.  (See Appendix \ref{sec:dsanp} for
details.)  Thus he proves that given a \sdsa instance with uniform load,
determining if $\opt=L$ is NP-complete, even if durations are restricted to
the set $\{1,2\}$, and by the above identity, the same is true for
\odspname.

\begin{figure}
\centerline{\includegraphics{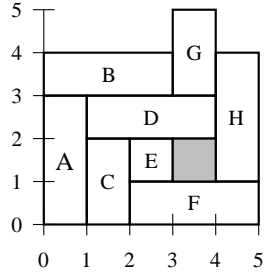}}
\caption{(From Buchsbaum et al.~\protect\cite{dsa:b+04}.) A set of sensors (in the
  form $(\ell(\cdot),r(\cdot),d(\cdot))$) $A=(0,1,3)$, $B=(0,3,1)$,
  $C=(1,2,2)$, $D=(1,4,1)$, $E=(2,3,1)$, $F=(2,5,1)$, $G=(3,4,2)$, and
  $H=(4,5,3)$.  The shaded region is a gap.  In this example, $L=4$ but
  $\opt=3$, which can be realized by sliding $G$ down so that $t(G)=1$.}
\label{fig:threefour}
\end{figure}

To establish a gap between $\opt$ and $L$, consider the example in Figure
\ref{fig:threefour}, in which $L = 4$ but $\opt = 3$.  
Scaling the durations shows that no approximation algorithm can
guarantee a ratio of better than $4/3$ with respect to $L$.

\subsection{A Dynamic Programming Solution for Small $L$}
\label{sec:dynam-progr-solut}
We give a dynamic program to answer the question: Is there a schedule $S$
such that $M(S)=T$ for a fixed $T$?  
\ignore{%
Assume first there exists such a schedule.  Then at any
$x$-coordinate $i$, the set of sensors that are live at some $i'<i$ must
have a schedule of duration $T$ when considered only up to $i$.  This is
the principle of optimality we shall use as we progress left-to-right.  
}
In
the following, we ignore portions of sensors that extend above level $T$ in
any schedule.

Define $\CS_{\le i} = \bigcup_{0\leq k \le i} \CS_k$.  Consider some
schedules $S_{i-1}$ of $\CS_{i-1}$ and $S_i$ of $\CS_i$ such that
$M(S_{i-1},i-1) = M(S_i,i)=T$.  We say that $S_{i-1}$ and $S_i$ are {\em
  compatible} if (i) for all $s\in S_{i-1} \cap S_i,\ t_{S_{i-1}}(s) =
t_{S_i}(s)$; and (ii) for all $j\in[0,T)$, $(i,j)$ is covered by $S_{i-1}$
or $S_i$.  The first condition stipulates that any sensor in both schedules
must have the same start time in each; the second requires
a sensor in $S_i$ to be scheduled to cover each level at which coverage
stops at $i-1$ in $S_{i-1}$.  
For each $i$, we populate an array $C_i$ indexed by
possible schedules of $\CS_i$.  For any $S_i$, define
$C_i[S_i]=1$ if there is a schedule $S$ of $\CS_{\le i}$ that respects
$S_i$ and has $M(S,x)=T$ for $0\leq x\leq i$; and $C_i[S_i]=0$ otherwise.
Then $C_i[S_i] = 1$ if and only if $M(S_i,i)=T$ and there exists some
schedule $S_{i-1}$ of $\CS_{i-1}$ such that $C_{i-1}[S_{i-1}]=1$ and
$S_{i-1}$ is compatible with $S_i$.  For $i=0$, $C_0[S_0]=1$ for precisely
those schedules $S_0$ of $\CS_0$ that have $M(S_0,0)=T$.  The dynamic
program then populates the arrays $C_i$ in increasing order of $i$, by
checking all schedules of $\CS_i$ for each $i$.  Ultimately we check if
there is some schedule $S_{2n-1}$ of $\CS_{2n-1}$ such that
$C_{2n-1}[S_{2n-1}]=1$.

\newpar{First Analysis.}
For a schedule $S_i$ of $\CS_i$, consider the union of the rectangles of
$S_i$, and denote by $\partial(S_i)$ the vertical boundaries of this union.
If $S_i$ is part of a minimal schedule $S$ of duration $T$, then
any rectangle of $S_i$ must cover some point on
$\partial(S_i)$ that is covered by no other rectangles in $S_i$.  
Thus $|S_i| \le 2T$, because $\partial(S_i)$ has total length
$2T$.  

Now we can analyze the dynamic program, which we restrict to consider only
minimal schedules.  The number of sensors in any schedule of $\CS_i$ is at
most $2T$, so there are at most $\binom{n}{2T}T^{2T}$ possible schedules of
$\CS_i$, as each potential set of $\eta$ sensors can be scheduled in
$T^\eta$ ways.  Each schedule of $\CS_i$ must be checked for compatibility
against each schedule of $\CS_{i-1}$, and checking compatibility of a pair
of schedules takes $O(T)$ time.  Hence the time to run the whole dynamic
program is $2n \left( \binom{n}{2T} T^{2T}\right)^2 O(T) = (nT)^{O(T)} =
(nL)^{O(L)} $.
To determine $\opt$, we run the dynamic program for each of the $L$
possible values of $T$, which does not affect the overall asymptotics.

\newpar{Partitioning the Dynamic Program.}
Now we restrict the $x$-coordinates on which we have to run the
dynamic program to those with relatively few live sensors.
Let $X = \{ i : |\CS_i| < 5T \}$.  
We claim that $\CS$ has a schedule of duration $T$ if and only if $\CS$ has
a schedule $S$ such that $M(S,i) \ge T$ for any $i\in X$.  We prove
the ``if'' part; the ``only if'' part is clear.

Assume  that there is a minimal schedule $S$ of duration $T$
that only covers $X$.  We show how to schedule the sensors not used
in $S$ to cover all $x$-coordinates.  Consider any maximal
interval $\bar{X}$ of $x$-coordinates not in $X$.  At most $4T$ sensors
from $S$ are live at any $i\in\bar{X}$, because any such sensor is also
live at either $\min(\bar{X})-1$ or $\max(\bar{X})+1$, and at most $2T$ are
live at either one.  By construction, there are at least $5T$ sensors live
at any $i\in\bar{X}$, so there are at least $5T-4T=T$ sensors live at $i$
that are not used by $S$ and hence are available, which suffice to cover
all the levels at $i$.  If such a sensor $s$ should also be live at another
$i'\in\bar{X}$ (or another $i'$ in another $\bar{X'}$), it reduces by one
both the number of potential uncovered levels and the number of available
sensors live at $i'$, so enough sensors will remain at $i'$.

Therefore we need only run the dynamic program on the $x$-coordinates in
$X$.  This takes only $2n\cdot T^{O(T)}$ time, because there are fewer than
$5T$ sensors live at any $i \in X$.  Thus we have:

\begin{theorem}
\label{thm:dp1}
\odspname~can be solved exactly in time $2n\cdot L^{O(L)}$.
\end{theorem}

\begin{corollary}
  \label{cor:dpc}
  \odspname~can be solved exactly in $\poly(n)$ time if $L< c \cdot \log
  n/\log\log n$ for some constant $c$ small enough.
\end{corollary}

Using a standard trick, a PTAS follows directly by truncating durations
appropriately. 

\begin{corollary}
  \label{thm:dp2}
  There is a PTAS for \odspname if $L < c \cdot \hmin \log n / \log\log n$
  for some sufficiently small constant $c$.
\end{corollary}


\subsection{Approximation Algorithms via Grouping}
\label{sec:appr-algor-via}

In this section, we give approximation algorithms via the {\em grouping}
technique, which is similar to the {\em boxing} technique of Buchsbaum et al.~\cite{dsa:b+04}.  
We know that the load $L$ is a natural upper bound on
$\opt$, and $L=\opt$ when all sensors have the same duration.  The basic
idea of grouping is to group shorter sensors into longer, virtual sensors
until all the sensors have equal duration, at which point the greedy
algorithm is invoked.  Essentially we must ensure that the load does not
decrease too much during the process, which is the central component of our
algorithms.

\newpar{Grouping Sensors.}
A {\em grouping} of a set $Y$ of sensors into a set of groups $G$ is a
partition of $Y$ into $|G|$ subsets, each of which is then replaced by a
rectangle that can be covered by the sensors in the group.  The {\em
  duration} of a group is defined to be the duration of the rectangle that
replaces it.  That is, these rectangles can be viewed as sensors in a
modified instance.  Then $L_G$ (rsp., $L_G(i)$) is defined to be the {\em
  load of the groups} (rsp., {\em at $i$}).  Note that $L_G(i) \le L_Y(i)$,
since portions of the sensors in a group that are overlapped or outside the
rectangle are not counted in $L_G(i)$.  In the following, we give
procedures to group a set $Y$ of sensors of unit duration into $G$ such
that $L_G(i)$ is not much smaller than $L_Y(i)$ for any $i$.  All of the
grouping procedures in this section run in polynomial time.

First, we give a grouping of a set of sensors that are all live at a fixed
$x$-coordinate.

\begin{lemma}
\label{lem:fixed}
Given a set $Y$ of unit-duration sensors, all live at some fixed
$x$-coordinate $x_0$, an integer group-duration parameter $D$, and a
sufficiently small positive $\epsilon$, there is a set $G$ of groups, each
of duration $D$, such that for any $i$,
\[ L_G(i) >  L_Y(i) / (1+\eps) - 4D \lceil 1 / \epsilon \rceil. \] 
\end{lemma}
\begin{proof}
  It is convenient to view a sensor $s$ as a point
  $(\ell(s),r(s))$ in the plane.  Note that all sensors live at $x_0$ are
  inside the rectangle $R_{x_0} = \{(x,y) : x\le x_0\le y\}$
  (Figures~\ref{fig:grouping}(a)--(b)).  First we partition the sensors of
  $Y$ into strips by repeating the following as long as sensors remain.

\begin{figure}[t]
\begin{center}
\centerline{\includegraphics{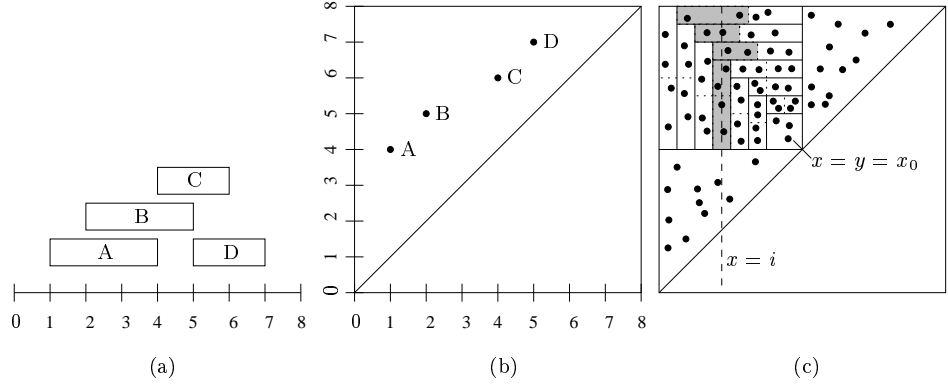}}
\caption{(From Buchsbaum et al.~\protect\cite{dsa:b+04}.) (a) Four sensors. (b) The sensors of (a)
  viewed as $(x,y)$ points. (c) Grouping a set of sensors with $D=2$ and
  $\eps = 1/2$.  The rectangle $R_i$ contains the set $Y$ of sensors.  The
  sensors are first partitioned into alternating vertical and horizontal
  strips of $D\lceil 1/\eps \rceil = 4$ each.  Within each strip, the
  sensors are grouped (dotted lines) into groups of $D=2$.  The groups that
  intersect the line $x=i$ are shaded.}
\label{fig:grouping}
\end{center}
\end{figure}
  
(1) Create a vertical strip containing the at most $D\lceil 1/\eps \rceil$ sensors
that remain with the smallest $\ell(\cdot)$ values.

(2) Create a horizontal strip containing the at most $D\lceil 1/\eps \rceil$
sensors that remain with the largest $r(\cdot)$ values.
  
Now for every vertical strip of $Y$, take the sensors in order of
decreasing $r(\cdot)$ value in groups of size $D$ 
(we may discard the last $<D$ sensors in the last strip). 
Similarly, for every horizontal strip,
take the sensors in order of increasing $\ell(\cdot)$ value in groups of
size $D$ (we may discard the last $<D$ sensors in the last strip).  
Replace each group $X$ with a larger rectangle   $s_X$ with
$\ell(s_X) = \max_{s\in X} \ell(s)$, $r(s_X) = \min_{s \in X} r(s)$, and
 $d(s_X) = \sum_{s\in X} d(s) = D$.
  
Consider any $i \le x_0$ (the case $i> x_0$ is symmetric), and examine
Figure~\ref{fig:grouping}(c).  All sensors live at $i$ are inside the
rectangle $R_i = \{(x,y) : x\le i \le y\}$.  Assume that the line $x=i$
intersects $k$ horizontal strips; then $R_i$ entirely contains at least
$k-1$ vertical strips, so $L_Y(i) \ge (k-1) D \lceil 1/\epsilon \rceil$.
For any group completely inside $R_i$, it contributes $D$ to both $L_Y(i)$
and $L_G(i)$; for any group completely outside $R_i$, it does not
contribute anything to either $L_Y(i)$ or $L_G(i)$.  So only the groups in
the $k$ horizontal strips and the single vertical strip intersected by the
line $x=i$ contribute to the difference, that is, $L_Y(i) - L_G(i) < kD +
D \lceil 1/\epsilon \rceil + D$, where the last term accounts for the fewer
than $D$ sensors that we did not group in the last strip.  Therefore,
  \[ L_G(i) > L_Y(i) - (k-1)D - (2+\lceil 1/\epsilon \rceil) D \ge (1 -
  \epsilon) L_Y(i) - 2D \lceil 1 / \epsilon \rceil \ge \frac{1}{1+2\eps}
  L_Y(i) - 2D \lceil 1 / \epsilon \rceil, \]
  for any $\epsilon \le 1/2$.  Replacing $\eps$ with $\eps/2$ gives the
  desired result.
\end{proof}

Next we use Lemma~\ref{lem:fixed} to group all sensors of unit duration.  

\begin{lemma}
  \label{lem:group}
  Given a set $Z$ of unit-duration sensors, an integer group-duration
  parameter $D$, and a sufficiently small positive $\epsilon$, there is a
  set $G$ of groups, each of duration $D$, such that at any $x$-coordinate
  $i$,
  \[ L_G(i) > L_Z(i)/(1 + \epsilon) - O(\log n \cdot D/\epsilon). \]
\end{lemma}
\begin{proof}
  Build an interval tree $\CT$ on
  the $x$-projections of the rectangles of $Z$.  For each node $v$ of $\CT$,
  let $Z_v$ be the set of sensors associated with $v$.  All the sensors of
  $Z_v$ are live at a fixed $x$-coordinate, namely the dividing line at
  $v$, and thus we can apply Lemma~\ref{lem:fixed} to $Z_v$ for each $v$.
  
  Consider any $x$-coordinate $i$.  For any two different nodes $u,v$ of
  $\CT$ at the same level, the sensors in $Z_u$ and the sensors in $Z_v$ do
  not overlap.  Thus the sensors live at $i$ are distributed to at most
  $O(\log n)$ nodes in $\CT$, because the interval tree has height $O(\log
  n)$.  By Lemma~\ref{lem:fixed}, we have
  $ L_G(i) > \sum_{v\in \CT} ( L_{Z_v}(i) / (1 + \epsilon) - 4 D \lceil
  1/\epsilon \rceil) =  L_Z(i) / (1 + \epsilon)- O(\log n \cdot
  D/\epsilon)$.
\end{proof}

\begin{figure}[t]
\centerline{\includegraphics{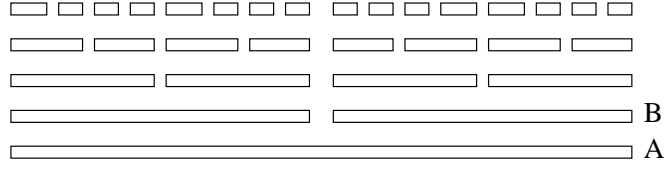}}
\caption{A bad example}
\label{fig:bad}
\end{figure}

\newpar{Remark.} The $O(\log n)$ factor in the error term of
Lemma~\ref{lem:group} cannot be removed.  Consider grouping the example in
Figure~\ref{fig:bad} with $D=2$.  First, at least half of sensor A has to
be wasted, because it is either grouped with some sensor in the left half
or some sensor in the right half.  Assume it is grouped with some sensor
in the left half.  Then by a similar argument, sensor B is cut in half, and
one of the halves has to be wasted.  Ultimately, we can find an
$x$-coordinate $i$ without a single group covering it; i.e., $L_G(i) = 0$,
but $L_Z(i) = \Omega(\log n)$.

\newpar{The Algorithm.}
\label{sec:alg1}
Let $\epsilon$ be a sufficiently small error parameter, and let $D=\hmax
\lceil 1/\epsilon \rceil$.  

\begin{algorithm}
  \begin{algorithmic}
\caption{Approximation algorithm via grouping}
\label{alg:1}
\STATE (1) Truncate each sensor of duration $d$ to $\lceil
(1+\epsilon)^k \rceil$, where $(1+\epsilon)^k \le d < (1+\epsilon)^{k+1}$
for some integer $k$.  Let $X$ be the set of truncated sensors.

\STATE (2) For each $d = \lceil (1+\epsilon)^k \rceil, k = \lfloor
\log_{1+\epsilon} \hmin \rfloor, \dots, \lceil \log_{1+\epsilon} \hmax - 1
\rceil$, do the following.  Let $X_d$ denote the set of truncated sensors
of duration $d$.  Scale each sensor in $X_d$ down by a factor of $d$, apply
Lemma~\ref{lem:group} with group-duration parameter $\lceil D/d \rceil$ and
the given $\epsilon$, and then scale the obtained groups back up by $d$.

\STATE (3) Let $G$ be the set of rectangles obtained from Step (2).
Truncate them so that they all have duration exactly $D$.  Call the
resulting set of rectangles $G'$.

\STATE (4) Apply the greedy algorithm to $G'$.
  \end{algorithmic}
\end{algorithm}

\begin{theorem}
  \label{thm:alg1}
  For any sufficiently small positive $\eps$, Algorithm~\ref{alg:1} runs in
  $\poly(n,1/\eps)$ time and gives a schedule of the \odspname~problem with
  duration at least $L/ (1 + \epsilon) - O\left(\hmax\log n \cdot 1/\eps^3
    \log (\hmax/\hmin) \right)$.
\end{theorem}
\begin{proof}  
  We will show that the truncating and grouping do not decrease the load at
  any $i$ excessively.

  By Lemma~\ref{lem:group}, Step 2 produces a grouping $G_d$ of $X_d$ of
  duration $\lceil D/d \rceil d$ such that at any $i$, $ L_{G_d}(i) >
  L_{X_d}(i)/ (1 + \epsilon) - O(\log n \cdot D / \epsilon)$.  Summing over
  all $d$, we have
\[
 L_{G}(i) > \frac{1}{1 + \epsilon} L_X(i) - O\left(\frac{D \log n \log
    (\hmax / 
    \hmin)}{\epsilon \log (1+\epsilon)} \right) 
= \frac{1}{1 + \epsilon} L_X(i) -
    O\left(\frac{\hmax\log(\hmax/\hmin) \log n}{\epsilon^3}
    \right).
\]
    
Truncating the sensors in Step (1) decreases their durations by at most
$1+\epsilon$, so $L_X(i) \ge \frac{1}{1+\epsilon} L(i)$.  Truncating
the groups in Step (3) decreases their durations by a factor of at most
$\frac{\lceil D / d \rceil d} {D} \le \frac{D+d}{D} \le 1 + \epsilon$, too.
Since $\frac{1}{(1+\epsilon)^3} \ge \frac{1}{1+7\epsilon}$, we have $
L_{G'}(i) > L(i)/(1+7\eps) - O\left(\hmax\log n \cdot 1/\eps^3
  \log(\hmax/\hmin)\right)$.  Finally, applying the greedy algorithm in
Step (4) yields a schedule of duration $ \min_i L_{G'}(i) > L
/(1+7\epsilon) - O(\hmax\log n \cdot 1/\eps^3 \log(\hmax/\hmin))$.
Replacing $\eps$ with $\eps/7$ gives the desired result.
\end{proof}

\begin{corollary}
  \label{cor:bigL1}
  There is a constant $c$ such that for any small enough positive
  $\epsilon$, the algorithm gives a schedule of duration at least $L /(1 +
  \epsilon)$ for any $L \ge c \cdot \hmax \log n \cdot 1 /\eps^4
  \log(\hmax/\hmin)$.
\end{corollary}

\newpar{An Alternative Algorithm.}
By bootstrapping Steps (1)--(3) of Algorithm \ref{alg:1}, we can replace
the $O(\log(\hmax / \hmin))$ factor with $O(1/\epsilon)$, leading to the
following result.

\begin{theorem}
  \label{thm:alg2}
  For any sufficiently small positive $\eps$, there is an algorithm that
  runs in $\poly(n,1/\eps)$ time and gives a schedule to the
  \odspname~problem with duration at least $L/ (1 + \epsilon) -
  O\left(\hmax\log n \cdot 1/\eps^4 \right)$.
\end{theorem}
\begin{proof}
  We are going to apply Steps (1)--(3) of Algorithm \ref{alg:1} repeatedly,
  grouping the smaller sensors so as to increase $\hmin$ until
  $\log(\hmax/\hmin)$ becomes small enough that we can apply
  Theorem~\ref{thm:alg1} to the resulted rectangles.
  
  For ease of presentation, we assume that $1/\eps$ is an integer.  Let $r$
  denote the ratio $\hmax / \hmin$.  Assume first that $\log r \ge
  1/\epsilon$, and set $\mu = \epsilon / \log r$ and $D = \lceil \mu^4
  \hmax \rceil$.  Apply Steps (1)--(3) of Algorithm~\ref{alg:1} to $\CS_s$,
  the set of sensors of duration at most $d'_{\max} = \lceil \mu D \rceil$,
  with group duration $D$ and error parameter $\mu$.  This yields a set of
  rectangles $G_s$ of duration $D$ such that for any $i$,
  \begin{eqnarray*}
    L_{G_s}(i) &>& \frac{1}{1+\mu} L_{\CS_s}(i) - O\left(\frac{d'_{\max}
        \log n \log(d'_{\max} / \hmin) }{\mu^3}\right) \\
    &>& \frac{1}{1+\mu} L_{\CS_s}(i) - O( \mu^2 \hmax  \log n \log (\mu^5 r))
    >  \frac{1}{1+\mu} L_{\CS_s}(i) - \frac{c_1 \epsilon^2 \hmax\log
      n}{\log r}.
~~~ \text{($c_1$ is some constant.)}
  \end{eqnarray*}

  Now consider $G_s$ as a set of sensors and the new problem instance $\CS'
  = G_s \cup (\CS \setminus \CS_s)$.  Its load at $i$ is 
\[ L_{\CS'}(i) >
  \frac{L(i)}{1+\mu} - \frac{c_1 \epsilon^2 \hmax \log n}{\log r}.\]
  Moreover, the new minimum duration of this problem instance is at least
  $\hmax'$, and the maximum duration remains $\hmax$, so the new ratio is $
  r' \le \frac{\hmax}{d'_{\max}} \le \frac{1}{\mu^5} = \frac{\log^5
    r}{\epsilon^5} \le \log^{10} r$, since $\log r \ge 1/\epsilon$.  For
  $\epsilon$ sufficiently small, we have $r' \le \sqrt{r}$; hence $\log r'
  \le \frac{1}{2} \log r$.

  Next repeat the procedure above, each time using new error parameter
  $\mu' = \epsilon / \log r'$, until it yields a problem instance $\CS^*$
  with minimum duration $d^*_{\min}$ for which $r^* = \hmax/\hmin^*$ is
  such that $\log r^* < 1/\epsilon$.  Let $r_0,\dots,r_k=r^*$ be the
  sequence of ratios and $L_0(i) = L(i), L_1(i), \dots, L_k(i)=L^*(i)$ be
  the sequence of loads.  We have
  \begin{eqnarray*}
    L^*(i) &>& \frac{1}{1+\eps/\log r_{k-1}} L_{k-1}(i) -
    \frac{c_1 \epsilon^2 \hmax \log n}{\log r_{k-1}} \\ 
    &>& \frac{1}{1+\eps/\log r_{k-1}}
    \left( \frac{1}{1+\eps/\log r_{k-2}} L_{k-2}(i) -
      \frac{c_1 \epsilon^2 \hmax \log n}{\log r_{k-2}} \right)
-  \frac{c_1 \epsilon^2 \hmax \log n}{\log r_{k-1}} \\
&\cdots & \\
&>& \left(\prod_{i=0}^{k-1} \left(1+\frac{\epsilon}{\log
    r_i}\right)\right)^{-1} 
    L(i) - 
\sum_{i=0}^{k-1}\left(\frac{1}{\log r_i}\right)c_1 \epsilon^2 \hmax \log
    n  \\
&\ge& \frac{1}{1+ c_2 \eps/\log r^*} L(i) - \frac{2}{\log r^*}
    c_1 \epsilon^2 \hmax \log n  \qquad \text{($c_2$ is some constant.)}\\
    &>& L(i)/(1 + 2c_2 \epsilon^2) - 4 c_1 \epsilon^3 \hmax \log n. 
  \end{eqnarray*}

  Let $L^* = \min_p L_{\CS^*}(i)$.  Finally, apply Theorem~\ref{thm:alg1}
  to $\CS^*$, which yields a schedule of duration at least
\[\frac{1}{1+\epsilon} L^* -
O\left(\frac{\hmax\log r^* \log n}{\epsilon^3} \right)
\ge \frac{1}{1+c_4\epsilon} L - O\left(\frac{\hmax \log n}{\epsilon^4}
\right), \]
for some constant $c_4$.  Replacing $\epsilon$ with $\epsilon / c_4$ gives
the desired result.
\end{proof}

\begin{corollary}
  \label{cor:bigL2}
  There is a constant $c$, such that for any small enough positive real
  $\epsilon$, the algorithm gives a schedule of duration at least
  $L/(1+\epsilon)$ for any $L \ge c \cdot \hmax \log n \cdot 1/\epsilon^5 $.
\end{corollary}


\subsection{An $O(\log\log n)$-Approximation for Arbitrary $L$}

Theorem~\ref{thm:alg2} yields a good approximation only when $d_{\max}$ is
small.  To extend this, we separate tall rectangles from the short
ones and handle the former individually.  Henceforth, we will analyze the
approximation ratio asymptotically, and we assume that all durations
are powers of 2, which at worst halves the duration of the schedule. 

Let $d_{\max} = 2^\rho$, and $\ell = \lceil 2\log\log n \rceil$.  We
partition $\CS$ into $\ell+1$ subsets $\CR_0, \dots, \CR_\ell$.  The first
subset $\CR_0$ consists of all sensors of duration at most $2^{\rho -
  \ell}$; then for each $\rho - \ell + 1 \le k \le \rho$, we put all
sensors of duration $2^k$ into one subset.  We call $\CR_0$ the {\em small
  subset} and the rest {\em large subsets}.  For any $x$-coordinate $i$,
we compute $L_{\CS_i \cap \CR_k}(i)$, the load of $\CS_i \cap \CR_k$ at
$i$, for $k=0, \dots, \ell$.  Let $m(i) = \argmax_k L_{\CS_i \cap
  \CR_k}(i)$; i.e., the sensors in $\CS_i \cap \CR_k$ have maximum load when
$k=m(i)$.  Break ties arbitrarily.  It is easy to see that the load of
$\CS_i \cap R_{m(i)}$ is at least $L/(\ell+1)$.

Next, for each $k$, we use $\CR_k$ to cover all the $x$-coordinates $i$
where $m(i) = k$, for a duration of $\Omega(L/\log\log n)$.  For a large
subset $\CR_k, k\ge 1$, because all its sensors have the same duration, we
can use the greedy algorithm to find a schedule of duration at least
$L/(\ell+1) = \Omega(L/\log\log n)$.  For the small subset $\CR_0$, we use
Theorem~\ref{thm:alg2} with $\eps = 1$.  Since the sensors in $\CR_0$ have
maximum duration $2^{\rho-\ell} < L / 2^\ell$, Theorem~\ref{thm:alg2} yields
a schedule of duration at least
\[
\frac{L}{2(\ell+1)} - O\left(\frac{L}{2^\ell} \log n\right) \ge
\Omega\left(\frac{L}{\log\log n}\right) - O\left(\frac{L}{\log^2 n} \log
  n\right) = \Omega\left(\frac{L}{\log\log n}\right).
\]

\begin{theorem}
  There exists a polynomial-time $O(\log\log n)$-approximation algorithm
  for the \odspname~problem. 
\end{theorem}



\section{Cube Cover}
\label{sec:rectangle}

\subsection{Hardness Results}
\label{ssec:rect-hardness}

When the $R(\cdot)$'s are axis-aligned rectangles 
and $U$ is a two-dimensional region, the problem is NP-hard even 
when the sensors have uniform duration, in contrast to the uniform-duration 
case for \odpname.  We use a reduction from an 
instance of \naesat with $n$ variables and $m$ clauses to an 
instance of \rectanglecover.  

An instance $I$ of \naesat
is a set $U = \{u_1, u_2, ..., u_n\}$ of variables
  and a collection $C = \{C_1, C_2, ..., C_m\}$ of clauses over $U$, such that
each clause $C_i \in C$ has $|C_i| = 3$.
The problem is to determine if
there a truth assignment for $U$ such that each clause in $C$ has at
least one true literal and at least one false literal \cite{GJ79}.  A key property of \naesat is that if $X$ is a satisfying assignment for an instance $I$ of \naesat, $\bar{X}$ is also a satisfying assignment of $I$.
 
\begin{figure}[tb]
\begin{minipage}[t]{0.33\textwidth}
\begin{center}
\includegraphics[width = 2 in]{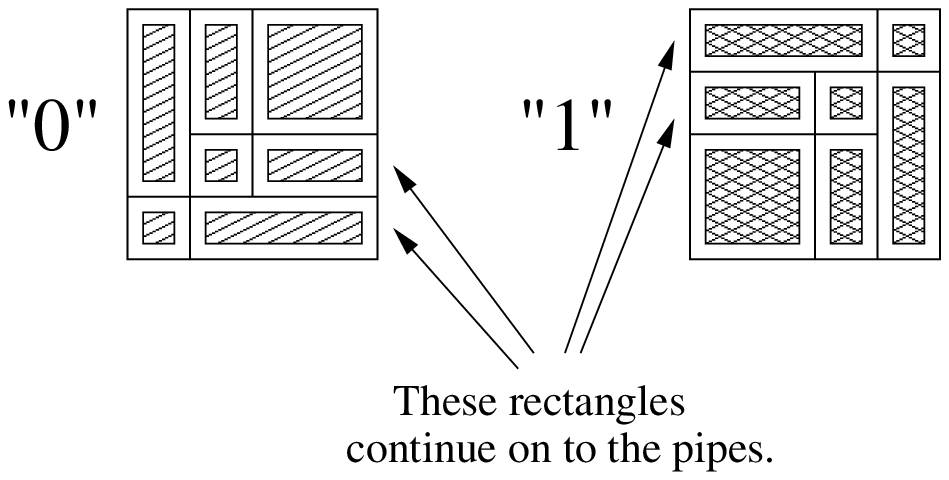}
\end{center}
\caption{Variables} \label{variable}
\end{minipage}%
\begin{minipage}[t]{0.33\textwidth}
\begin{center}
\includegraphics[width = 2 in]{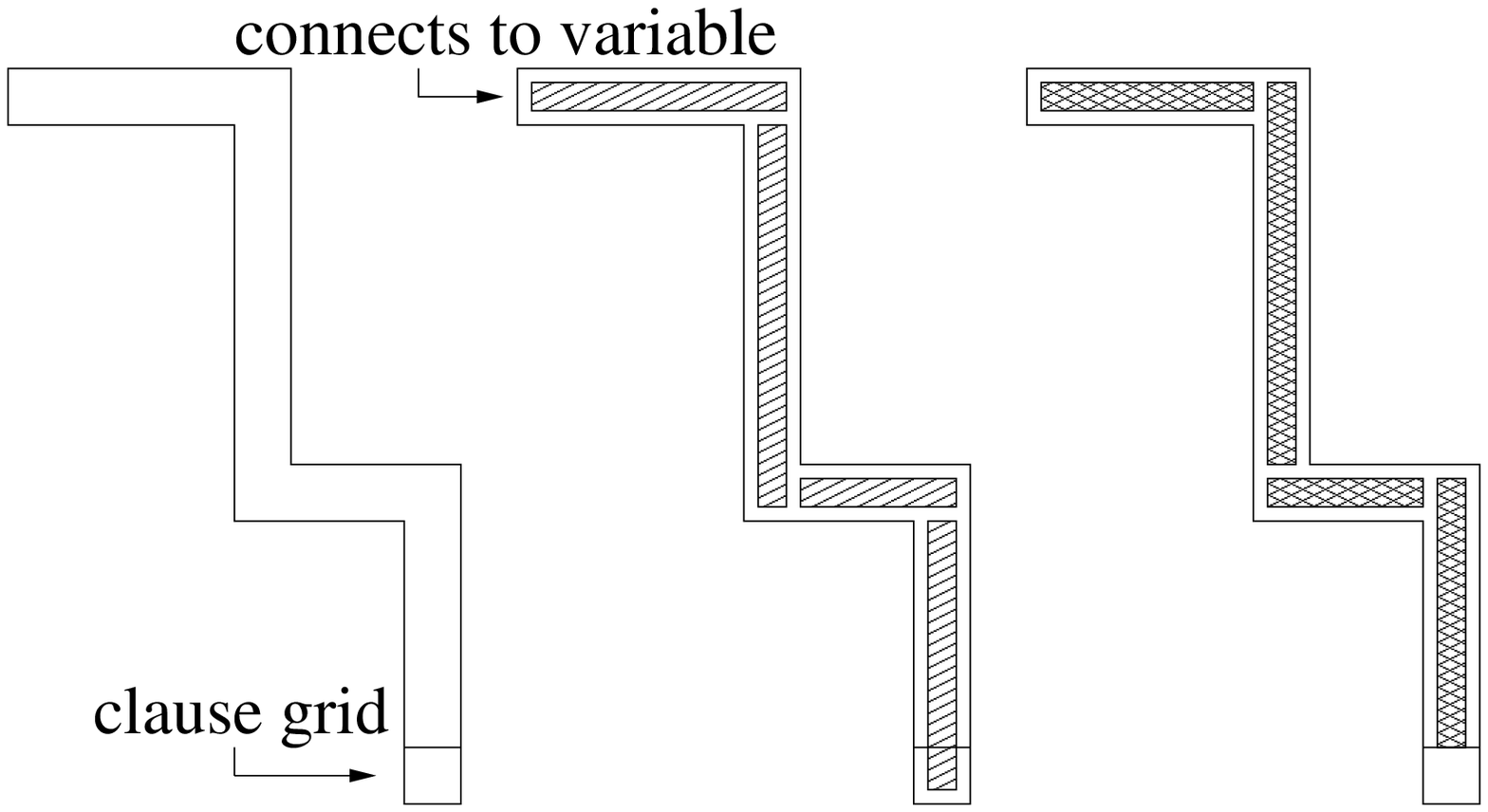}
\end{center}
\caption{Pipes} \label{pipe}
\end{minipage}%
\begin{minipage}[t]{0.33\textwidth}
\begin{center}
\includegraphics[width = 2 in]{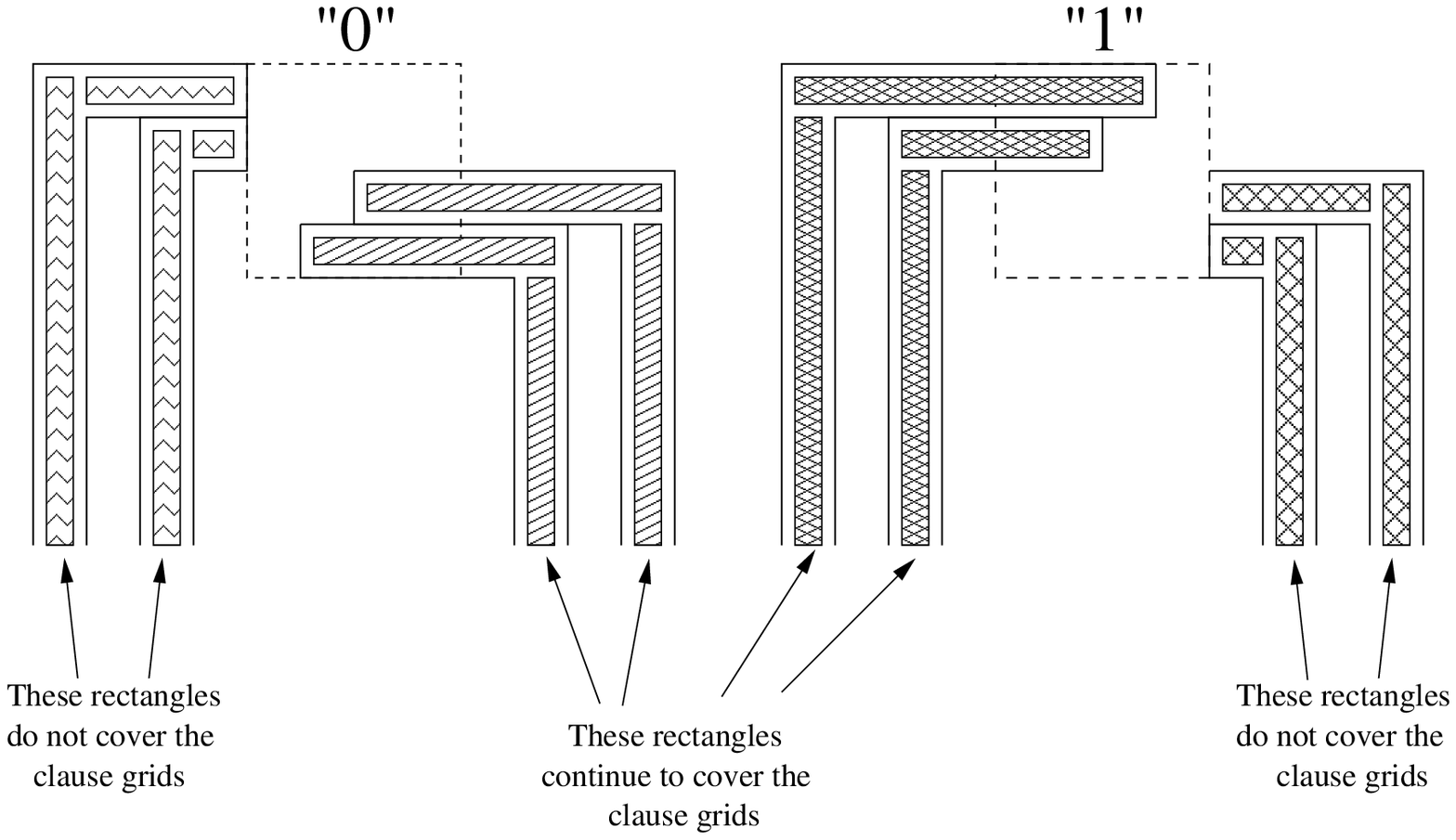}
\end{center}
\caption{Variables and Pipes} \label{varpipes}
\end{minipage}
\end{figure}

  Given $I$, we construct an associated graph $G(I)$, with vertices for each variable 
  and each clause.  We draw and edge between a clause vertex and a variable vertex if the variable appears in the clause.  The graph is drawn on a planar grid within a bounding box $U$.  From $G(I)$, we construct an instance $\CS(I)$ of \rectanglecover that has a schedule of duration 2 if and only if $I$ is satisfiable. If $I$ is unsatisfiable, $\CS(I)$ has a schedule of duration 1.  We describe the construction in more detail below:

\begin{description}
\item[Variables.]

  Each variable will be represented by a collection of
  rectangles that cover a square grid.   Each rectangle 
  covering the variable gadget has unit duration. The rectangles 
  that cover the variable gadget are shown in
  Figure~\ref{variable}, arranged in the ``true'' and ``false'' encodings.  
  Using a mixture of the rectangles in the ``true'' and ``false'' encodings leads to
  a suboptimal schedule; such a mixture is called an {\em improper cover}. 

\item[Pipes.]

  We connect each variable to each clause that contains it via \emph{pipes}.  The pipes
  are drawn using polygonal lines in the plane.  A pipe
  corresponding to a positive (rsp., negative) occurrence of a variable in a clause leaves
  via the left (rsp., right) side of the variable gadget. All rectangles that cover a 
  pipe have unit duration.  A pipe, as it is drawn within the bounding box, is 
  the leftmost image in Figure~\ref{pipe}. To the right of the pipe are the two proper
  configurations for covering the pipe.  The first of the configurations shows 
  how the pipe will be covered when the clause to which it connects is satisfied by 
  the variable assignment.  We refer to this as an ``on'' signal.  The second configuration 
  shows how the pipe is covered otherwise.  We refer to this as an ``off'' signal.  
  Using a mixture of rectangles in the two configurations (or an improper cover) 
  will lead to a suboptimal schedule.  Figure~\ref{varpipes} shows how
  the variables and pipes connect.

\item[Clauses.]

A clause is represented by a square grid at the end of a pipe. A clause is covered if a 
variable contained in it is satisfied.  All other parts of the 
bounding box are covered with two of the same rectangles, each with unit duration so we can 
guarantee that every point in $U$, not part of the variable, pipe, or clause gadgets, can be 
covered for a total of 2 time units. 
\end{description}

In the worst case, each of the variables require a square grid of size
$O(m)$, hence the entire grid must contain $O(nm)$ rows and
$O(nm)$ columns. Our construction requires $O(n^2m^2)$ unit squares 
and thus can be performed in polynomial time and space in the size of the input.

\begin{proposition} \label{suboptimal}
A variable gadget can be covered for two time units using proper
covers. If an improper cover is used for $d$ time units, then the gadget
can only be covered for $2-d$ time units.
\end{proposition}
\begin{proof}
Each point in the variable gadget is covered by exactly two rectangles. If
proper covers are used, the bounding rectangle can be covered for two
units (one for each cover). Note that this is the only way to partition
the rectangles into disjoint feasible covers.

Suppose an improper cover is used for $d$ units and the two proper covers are
used for $x, y$ time units. Thus the total time for which this rectangle
is covered is $x+y+d$. Since at least one rectangle from each
proper cover must be in an improper cover, $x+d\le 1$ and $y+d\le 1$, and
therefore $x+y+d \le 2-d$.
\end{proof}

\begin{proposition} \label{clause}
The clause square is covered if a variable contained
in it is satisfied.
\end{proposition}
\begin{proof}
If clause $C$ contains variable $u$ and variable $u$ is satisfied
in that clause, this means the configuration shown in the middle image
of Figure~\ref{pipe} is chosen so the clause grid is covered.
\end{proof}

\begin{proposition} \label{complement}
If $X$ is a satisfying assignment for \naesat, then
$\bar{X}$ is a satisfying assignment as well.
\end{proposition}
\begin{proof}
If $X$ is a satisfying assignment for \naesat, then
every clause contains at least one variable, $u$, that is true
and one variable, $v$, that is false. The complement,
$\bar{X}$, makes $u$ false and $v$ true.  $\bar{X}$
ensures that every clause contains at least one true variable
and one false variable and thus is a satisfying assignment.
\end{proof}

\begin{lemma} \label{lem:duration2}
  If $I$ is satisfiable, then the instance of \rectanglecover has a schedule
  of duration 2.
\end{lemma}
\begin{proof}
  Let $X$ be a satisfying assignment of $I$. For each variable that is set
  to 1, choose the ``1'' orientation for the corresponding variable gadget
  and the corresponding ``on'' signal in the pipes connecting this
  variable to the clauses that contain a positive occurrence of the variable. 
  Choose the ``off'' signal in pipes connecting this variable to the clauses 
  that contain a negative occurrence of the variable.  We repeat this process 
  for variables set to ``0''.

  Since $X$ is a satisfying assignment, each clause grid will be covered. 
  Moreover, since $\bar{X}$ is also a satisfying
  assignment, we can repeat this process for another time unit.
  Note that each pipe is used exactly once to send a ``on'' signal and once to
  send a ``off'' signal, so it can be covered for two time units.  Likewise, each 
  variable is used exactly once as ``1'' and once as ``0'', so it can also be
  covered for two time units.
\end{proof}

\begin{lemma} \label{lem:duration1}
  If $I$ is unsatisfiable, then the instance of \rectanglecover has a
  schedule of duration at most 1.
\end{lemma}
\begin{proof}
If all variables are covered using proper covers, we obtain a valid
assignment, so some clause rectangle must remain uncovered (since $I$
is unsatisfiable). In order to cover this clause rectangle, we must use 
an improper cover.  Our construction requires that an improper cover creates
an overlap either in the variable gadget or the pipes.  Since the load at every 
point in the variable gadget and the pipes is 2 and a rectangle must be used 
in its entirety or not at all, the maximum schedule has duration 1.   
\end{proof}


\begin{theorem}
\label{thm:rcnp}
\rectanglecover is NP-hard and does not admit a PTAS, even with uniform duration.
\end{theorem}
\begin{proof}
NP-hardness follows from Lemma~\ref{lem:duration2} and Lemma~\ref{lem:duration1}.

  Assume we have a PTAS  $\CP$ for \rectanglecover.  On input $(\CS(I), \epsilon)$,
  where $\epsilon > 0$ and $\CS(I)$ is an instance of \rectanglecover
        induced by the above construction, $\CP$
  would output a solution with duration $T$, where $T \geq (1-\epsilon)
  \cdot OPT$.  Setting $\epsilon = 0.25$, $T\geq 1.5$ when $OPT = 2$, 
  and $0.75 \leq T\leq 1$ when $OPT = 1$.  Thus we can
  use $\CP$ to distinguish
  between satisfiable and unsatisfiable instances of \naesat.
\end{proof}

\subsection{Rectangles With Unit Duration}
\label{sec:disks-with-arbitrary}

We consider approximation algorithms for \tdpname if
all sensors have unit duration.  First we prove a technical lemma, which
actually holds for arbitrary sets.

\begin{lemma}
\label{lem:logU}
Let $U$ be finite set with $m$ elements. For each $s\in \CS$, $R(s)$ is an
arbitrary subset of $U$ with unit duration.  There exists some constant $c$
large enough, such that if $L>c \ln m$, then in polynomial time we can find
a subset $\CR \subseteq \CS$ and a schedule of $\CR$ with duration at least
$L/\ln m$, such that the remaining load $L_{\CS \setminus \CR} \ge L/2$.
\end{lemma}
\begin{proof}
  We take covers from $\CS$ one by one.  Let $L_i$ be the load of the
  remaining sensors after the $i\ith$ cover has been taken.  For the
  $(i+1)\ith$ cover $\CX$, we take each remaining sensor into $\CX$ with
  probability $p=c\ln m / L_i$.  Then we check if (1) $\CX$ is a valid
  cover, and (2) the remaining load $L_{i+1} > L_i - \frac{1}{2}\ln
  m$.  For any $x \in U$, the probability that $x$ is not covered is at
  most $(1-p)^{L_i} < m^{-c}$, so (1) occurs with probability at least
  $1- m^{1-c}$ (probability of union of events).  For any $x\in U$, the
  probability that $L_i(x) \le L_i - \frac{1}{2}\ln m$ is at most
  $m^{-(2c-1)^2/8c}$ (Chernoff bound), so (2) occurs with probability
  at least $1-m^{1-(2c-1)^2/8c}$.  Thus we can choose $c$ large enough so
  that both (1) and (2) occur with high probability (e.g., $>1/2$).  We repeatedly take $\CX$ until this happens and then  proceed
  to the next cover.  We repeat this procedure until $L_{i+1}$ drops below
  $L/2$, and the lemma follows.
\end{proof}

The basic idea of our algorithms is the following. Take a partition of $U$
with a small number of cells, and then crop $R(\cdot)$ so that each
sensor fully covers a number of cells but is completely disjoint from the
rest.  We ensure that the load does not decrease by more than a
constant factor and then apply
Lemma~\ref{lem:logU}.

\begin{theorem}
\label{thm:rectangle}
If each sensor $s \in \CS$ has unit duration, then there is a
polynomial-time $O(\log(n/L))$-approximation algorithm for \tdpname.
\end{theorem}
\begin{proof}
  We assume $L>c\ln n$ for some large constant $c$; otherwise we just take one
  cover, and the theorem follows.  It is well known that sets of rectangles
  in the plane admit $(1/r)$-cuttings; there exists a subset $\CR \subset
  \CS$ of $r\log r$ rectangles such that in the partition $\CA_{\CR}$
  determined by the rectangles of $\CR$, each face is intersected by the
  boundaries of at most $cn/r$ rectangles of $\CS$ \cite{bs-ca-95}. We
  choose $r = \lceil 2cn/L \rceil$, so $cn/r \le L/2$.

  Let $f$ be a face of $\CA_{\CR}$, and let $\CS_f\subseteq\CS$ denote the
  subset of rectangles that fully contain $f$. Since the load at every
  point in $f$ is at least $L$ and only $L/2$ rectangles partially cover
  $f$, we derive $|\CS_f|\geq L/2$.  Now replace each rectangle $R(s)$
  by a cropped region that consists of all faces of $\CA_{\CR}$
  that $s$ fully covers.  This yields an instance of \sensorcover, with a
  universe of size $r^2\log^2 r$ and load $L' \ge L/2$.  Applying
  Lemma~\ref{lem:logU} yields the desired result.
\end{proof}

An improved bound can be obtained when all the $R(\cdot)$'s have the same
size by a more careful cropping scheme.

\begin{theorem}
\label{thm:square}
If each sensor $s \in \CS$ has unit duration and each $R(s)$ is a unit square,
then there is a polynomial-time $O(\log(L_{\max}/L))$-approximation
algorithm for \tdpname, where $L_{\max} = \max_x L(x)$. Note that $L_{\max}
\le n$.
\end{theorem}
\begin{proof}

\def\td{L_{\max}}
\def\nmax{{n_{max}}}

We assume $L>c\ln L_{\max}$ for some large constant $c$; otherwise we just take
one cover, and the theorem follows.  We draw a unit-coordinate grid $\Gamma$
inside $U$.  There are only $O(n/L)$ cells in $\Gamma$.  For a cell
$\gamma\in \Gamma$, let $\CS(\gamma)\subseteq \CS $ denote the set of
squares of $\CS$ that intersect $\gamma$. Let $\nmax =
\max_\gamma|\CS(\gamma)|$.  Packing arguments imply $\nmax \leq 4\td$. 

Two cells in $\Gamma$ are {\em independent} if they are at least two grid
cells apart from each other in both dimensions.  It is easy to see that we
can partition $\Gamma$ into 9 {\em independents sets} $\Gamma_1, \dots,
\Gamma_9$, where all cells in any one set are mutually independent.  In
the following, we will show how to make $\Omega(L/\ln(\nmax/L))$ covers for
$\Gamma_1$, such that the remaining load is at least $L/8$.  Then we repeat
the process for $\Gamma_2, \dots, \Gamma_9$, and ultimately we derive a
schedule that covers all cells with duration $\Omega(L/\ln(\nmax/L)) =
\Omega(L/\ln(L_{\max}/L))$.

By the definition of independence, we can isolate the cells in $\Gamma_1$
and only need to show that for any $\gamma \in \Gamma_1$, we can make
$\Omega(L/\ln(\nmax/L))$ covers from $\CS(\gamma)$ without decreasing the
load of any of its neighboring 8 cells by more than a factor of 8.  Since
the load of $\CS(\gamma)$ inside $\gamma$ is at least $L$, following the
same approach as in the proof of Theorem~\ref{thm:rectangle}, we can build
a partition $\CA$ in $\gamma$ and its neighboring cells such that each face
of $\CA$ is intersected by the boundaries of at most $L/2$ squares from
$\CS(\gamma)$.  $\CA$ has size $r^2\log^2r$, where $r = \lceil 2c\nmax /L
\rceil$.  We further partition the faces of $\CA$ that are intersected by
the boundary of $\gamma$, such that each face of $\CA$ is either inside
$\gamma$ or outside.  This increases the size of $\CA$ by a factor at most
2.  Let $\CF$ be the set of faces of $\CA$ that are fully covered by at
least $L/4$ sensors from $\CS(\gamma)$.  $\CF$ includes all faces inside
$\gamma$ and some faces outside.  For any face not in $\CF$, the load of
$\CS \setminus \CS(\gamma)$ must be at least $L-L/2-L/4=L/4$, so we can
ignore it.  Consider the faces in $\CF$.  We crop the
squares of $\CS(\gamma)$ according to $\CF$ in the same way as in the proof
of Theorem~\ref{thm:rectangle}.  After cropping, by construction the load
at each face of $\CF$ is still at least $L/4$.  Then we apply
Lemma~\ref{lem:logU} with $U=\CF$ and $\CS(\gamma)$, which gives us
$\Omega(L/\ln(\nmax/L))$ covers while the remaining sensors have load at
least $L/8$ for any face of $\CA$.
\end{proof}

\nib{Remark.} These results can be extended to any collection of shapes
that admit small cuttings: disks, ellipses, etc.



\section{Sensor Cover}
\label{sec:kcover}

Now consider the general \aspname problem, in which each $R(\cdot)$ is an arbitrary
subset of a finite set $U$ of size $|U| = O(n)$.  We show that a random
schedule of the sensors yields an $O(\log n)$-approximation with high
probability.  This result extends that of Feige et al.~\cite{FHK00}, which
deals with the unit duration case.

Let $T = c L / \ln n$, where $c$ is some constant to be determined later.
We show that if we choose the start time of each sensor randomly between 0
and $T$, then we will have a valid schedule with high probability.  In
order to avoid fringe effects, we must choose positions near 0 or $T$ judiciously.
  More precisely, for a sensor $s$ of
duration $d(s) < T$, we choose its start time $t(s)$ uniformly at random
between $-d(s)$ and $T$; if $t(s) < 0$, we reset it to 0.  If $d(s) \ge T$,
we simply set $t(s)=0$.  Divide $T$ evenly into $2n$ time intervals
$[t_0=0, t_1], [t_1, t_2], \dots, [t_{2n-1}, t_{2n}=T]$, each of length
$T/2n$.  If $d(s) \ge T/n$, it is easy to see that for any $x \in R(s)$ and
in any time interval, $x$ is covered by $s$ with probability at least
$(d(s) - T/2n) / (T+d(s)) \ge \frac{1}{4} \cdot d(s)/T$.

Consider any $x \in U$, and let $\{s_1, \dots, s_k\}$ be the set of sensors
live at $x$ with durations at least $T/n$.  We know that $\sum_{i=1}^k
d(s_i) \ge L - T/n \cdot n \ge L/2$.  In any time interval $[t_i,
t_{i+1}]$, the probability that $x$ is not covered is at most
\[ \prod_{i=1}^k \left(1-\frac{d(s_i)}{4T}\right) \le \prod_{i=1}^k
\exp\left(-\frac{d(s_i)}{4T}\right) \le \exp\left(-\frac{L}{8T} \right) =
\exp\left(-\frac{\ln n}{8c}\right) = n^{-\frac{1}{8c}}. \] 
There are
only $O(n^2)$ different $(x, [t_i, t_{i+1}])$ pairs, so the
probability that some $x \in U$ is not covered at some time is at most
$O(n^2) \cdot n^{-\frac{1}{8c}} = O\left(n^{2-\frac{1}{8c}}\right)$.  
Choosing $c < 1/16$ yields a high probability of obtaining a valid
schedule.

The algorithm can be de-randomized using the method of conditional
probability.  We omit the details. 

It is not hard to see that \scp can be reduced to \sensorcover. Given an
optimal schedule produced by an algorithm for \sensorcover, we can
``snap'' each starting time $t(s)$ to the integer $\lceil t(s) \rceil$
without introducing any gaps or decreasing the total duration. Hence, the
lower bound of Feige \etal\cite{FHK00} applies.

\begin{theorem}
\label{thm:sensor-cover}
  There exists a polynomial-time $O(\log n)$-approximation algorithm
  for the \aspname problem. This bound is tight up to constant factors.
\end{theorem}


\section{Conclusions and Open Problems} \label{sec:open}

Many questions remain open.  Ideally we
would like to prove stronger hardness results or find better
approximation algorithms in order to narrow the gap between our lower
and upper bounds.  In fact, we have not ruled out the possibility of a
PTAS for the \intervalcover problem, although it cannot be in
terms of $L$.  It would be interesting to
see if other techniques for geometric optimization problems could be
applied to our problem as well.

We are also interested in understanding preemptive schedules better.
For \intervalcover, a simple algorithm based on maximum flow
yields an optimal preemptive schedule in polynomial time.
In higher dimensions, however,
it is not fully understood in which situations non-preemptive schedules
are sub-optimal when compared with the best preemptive schedules.  In general, we would like to uncover the
relationship between the load of the problem instance, the duration
of the optimal preemptive schedule, and the duration of the optimal non-preemptive
schedule.

\paragraph{Acknowledgement.}
We thank Nikhil Bansal for pointing us
to the paper by Kenyon and Remila \cite{strip:kr00}.

\bibliographystyle{plain}
\bibliography{main}

\begin{thebibliography}{1}

\bibitem{AGP04}
Z.~Abrams, A.~Goel, and S.~Plotkin.
\newblock Set {K}-cover algorithms for energy efficient monitoring in wireless
  sensor networks.
\newblock In {\em Proc. 3rd Int'l. Symp. Information Processing in Sensor
  Networks (IPSN)}, pages 424--432, 2004.

\bibitem{dsa:b+04}
A.~L. Buchsbaum, H.~Karloff, C.~Kenyon, N.~Reingold, and M.~Thorup.
\newblock {{\em OPT}} versus {{\em LOAD}} in dynamic storage allocation.
\newblock {\em SIAM J. Computing}, 33(3):632--46, 2004.

\bibitem{DVCS04}
S.~Dasika, S.~Vrudhula, K.~Chopra, and R.~Srinivasan.
\newblock A framework for battery-aware sensor management.
\newblock In {\em Proc. Design, Automation and Test in Europe Conf. and Expos.
  (DATE)}, pages 1--6, 2004.

\bibitem{bs-ca-95}
M.~de~Berg and O.~Schwarzkopf.
\newblock Cuttings and applications.
\newblock {\em Int'l. J. Comp. Geom. \& Appl.}, 5(4):343--355, 1995.

\bibitem{FHK00}
U.~Feige, M.~M. Halld{\'{o}}rsson, G.~Kortsarz, and A.~Srinivasan.
\newblock Approximating the domatic number.
\newblock {\em SIAM J. Computing}, 32(1):172--195, 2002.

\bibitem{GJ79}
M.~R. Garey and D.~S. Johnson.
\newblock {\em Computers and Intractability: {A} Guide to the Theory of
  {NP}-Completeness}.
\newblock W.H. Freeman and Company, 1979.

\bibitem{strip:kr00}
C.~Kenyon and E.~Remila.
\newblock A near-optimal solution to a two-dimensional cutting stock problem.
\newblock {\em Math. of Op. Res.}, 25(4):645--656, 2000.

\bibitem{PH03}
M.~Perillo and W.~Heinzelman.
\newblock Optimal sensor management under energy and reliability constraints.
\newblock In {\em Proc. IEEE Wireless Communications and Networking Conf.
  (WCNC)}, pages 1621--1626, 2003.

\bibitem{SP01}
S.~Slijepcevic and M.~Potkonjak.
\newblock Power efficient organization of wireless sensor networks.
\newblock In {\em Proc. IEEE Int'l. Conf. on Communications (ICC)}, pages
  472--476, June 2001.

\end{thebibliography}

\appendix

\section{NP-Completeness of Dynamic Storage Allocation}
\label{sec:dsanp}

The following proof was given by Larry Stockmeyer to David Johnson
and cited as the ``private communication'' behind the
NP-completeness result in Garey and Johnson \cite[Problem SR2]{GJ79}.
To our knowledge, this proof has not previously appeared;
we include it here essentially verbatim for the more specific results
needed in Section \ref{sec:icnp}.
All credit goes to Larry Stockmeyer.
We thank David Johnson for supplying the proof to us.

\smallskip

\noindent\dsa

\nib{Instance:}
Set $A$ of items to be stored,
each $a\in A$ having a size $s(a)$, an arrival time $r(a)$,
and a departure time $d(a)$ (with $d(a)>r(a)$),
and a positive integer storage size $D$.

\nib{Question:}
Is there an allocation of storage for $A$; i.e.,
a function $\sigma : A \rightarrow \{1,2,\ldots,D\}$
such that for every $a\in A$ the allocated storage internal
$I(a) = [\sigma(a),\sigma(a)+s(a)-1]$ is contained in $[1,D]$
and such that,
for all $a,a'\in A$ with $a\not= a'$, if $I(a)\cap I(a')$ is nonempty
then $[r(a),d(a)) \cap [r(a'),d(a'))$ is empty?

\nib{Theorem:}
\dsa is NP-complete, even when restricted to instances where $s(a)\in\{1,2\}$
for all $a$.

The reduction is from the \tpp problem.

\smallskip
\noindent\tpp

\nib{Instance:} Set $W$ of $3m$ elements, a bound $B$, and a positive integer
size $z(w)$ for each $w\in W$, such that $\sum_{w\in W} z(w)=mB$,
and $B/4<z(w)<B/2$ for all $w$.

\nib{Question:}
Can $W$ be partitioned into disjoint sets $W_1,\ldots,W_m$ such that,
for $1\leq i\leq m$, $\sum_{w\in W} z(w) = B$?

\tpp is strongly NP-complete \cite{GJ79}; i.e., there is a polynomial 
$p$ such that it is NP-complete when restricted to instances $I$ where $z(w)\leq p$
(the length of $I$) for all $w$.  The condition $B/4<z(w)<B/2$ is not used in the following reduction.

Given an instance of \tpp as above,
the corresponding instance of \dsa has storage size $D=m(B+1)+2$.
The instance is described by giving a time-ordered sequence of arrivals
and departures of items of various sizes.
It is also convenient for the description to allow an item $t$ of size 2
to arrive several times,
provided that it first departs before arriving again.
Thus, if this item $t$ arrives $k$ times in the entire description,
there are really $k$ different items, $t_i$ for $1\leq i\leq k$,
all of size 2, and no two of them exist at the same time.

In the following, the items $f_i$, $g_i$, and $h_i$ all have size 1.
Begin by having $D$ items $f_1,\ldots,f_D$
arrive.  Next $f_1$ and $f_2$ depart, then $t$ arrives and departs,
then $g_1$ and $g_2$ arrive.
Now do in sequence for $i=2,3,\ldots,D-1$:
\begin{enumerate}
\item items $g_i$ and $f_{i+1}$ depart;
\item then item $t$ arrives and departs;
\item then items $h_i$ and $g_{i+1}$ arrive.
\end{enumerate}
Finally, $g_1$ departs and $h_1$ arrives,
and then $g_D$ departs and $h_D$ arrives.
At this point, it can be seen that the order of the items in storage must
be $h_1,h_2,\ldots,h_{D-2},h_{D-1},h_D$ or
$h_1,h_2,\ldots,h_{D-2},h_D,h_{D-1}$ or the reversal of one of
those orderings.

Now $h_i$ departs for every $i\leq D-2$
that is not a multiple of $B+1$.
At this point, storage consists of $m$ blocks of free space,
each block has length $B$,
and there are barriers (namely, $h_{B+1}$, $h_{2B+2}$, $\ldots$)
between each pair of adjacent blocks.

Consider first the case that sizes of items are not restricted to 1 and 2.
For each $w\in W$, an item $q_w$ of size $z(w)$ arrives.
If the \tpp instance has a solution $W_1,\ldots,W_m$,
then the items $q_w$ with $w\in W_1$ can go into the first block
of free space, the items $q_w$ with $w\in W_2$ can go
into the second block of free space, and so on for $W_3,\ldots,W_m$.

Conversely, if all the $q_w$ fit into the $m$ free blocks of length $B$, then
there must be a solution to the \tpp instance.

Consider now the case that sizes of items are restricted to 1 and 2.
For each $w\in W$, the item $q_w$ of size $z(w)$ is replace by $z(w)$
items $q_{w,i}$ for $1\leq i\leq z(w)$, all of size 1.
For each $w$, additional arrivals and departures are now added to ensure that the $z(w)$
items $q_{w,i}$ were placed in the same block of free space.
The method is similar to the one used above with
$f$'s, $g$'s, and $h$'s.
First $q_{w,1}$ and $q_{w,2}$ depart,
then $t$ arrives and departs,
then $q'_{w,1}$ and $q'_{w,2}$ arrive.
The for $i=2,3,\ldots,z(w)-1$:
$q'_{w,i}$ and $q_{w,i+1}$ depart,
then $t$ arrives and departs,
then $q''_{w,i}$ and $q'_{w,i+1}$ arrive.
If the $q_{w,i}$ were not placed in the same free block,
then at some point the two units of free space 
formed by the departure of two items must be separated
by a barrier,
so the item $t$ of size 2 cannot be allocated.

This is a polynomial-time transformation,
because the numbers $z(w)$ are bounded above by a polynomial 
in the length of the \tpp instance.
Note further that the reduction
constructs a \dsa instance of uniform
load with the ultimate determination
being whether $\opt=L$.


\end{document}